\documentclass{article}
\usepackage{spconf,amsmath,graphicx}
\usepackage{booktabs} 
\usepackage{url}

\title{Clean Label Attacks against SLU Systems}
\name{Henry Li Xinyuan, Sonal Joshi, Thomas Thebaud, Jesus Villalba, Najim Dehak, Sanjeev Khudanpur}
\address{
  $^1$Center for Language and Speech Processing, Johns Hopkins University, USA
}
\begin{document}
%
\maketitle
\begin{abstract}

Poisoning backdoor attacks involve an adversary manipulating the training data to induce certain behaviors in the victim model by inserting a trigger in the signal at inference time. We adapted clean label backdoor (CLBD)-data poisoning attacks, which do not modify the training labels, on state-of-the-art speech recognition models that support/perform a Spoken Language Understanding task, achieving 99.8\% attack success rate by poisoning 10\% of the training data. We analyzed how varying the signal-strength of the poison, percent of samples poisoned, and choice of trigger impact the attack. We also found that CLBD attacks are most successful when applied to training samples that are inherently hard for a proxy model. Using this strategy, we achieved an attack success rate of 99.3\% by poisoning a meager 1.5\% of the training data. Finally, we applied two previously developed defenses against gradient-based attacks, and found that they attain mixed success against poisoning.
\end{abstract}
\section{Introduction}


Data poisoning attacks \cite{Big,trojaning} have long been shown to be effective against neural networks for modalities including images \cite{Badnets,CV,Badtrack}, text \cite{LLM}, and audio \cite{audio}. 
Poisoning attacks entail an adversary manipulating a subset of the training data in order to induce a certain behavior in the victim model at inference time, for example, performance degradation or backdoor access.
These attacks have emerged as a realistic and effective threat model as the training of large models \cite{whisper-style,gpt-3} increasingly necessitates sourcing large amounts of data from the Internet, some of which might be vulnerable to manipulation by an adversary \cite{practical}. 
Poisoning attacks can be broadly classified into two types--Dirty Label Poisoning  (modify both data and labels) and Clean Label Poisoning (modify data but not labels). 
Comprehensive studies have been performed on the efficacy of poisoning attacks against classification models, and successful dirty label poisoning attacks have been crafted against sequence tasks like machine translation \cite{blackbox-nmt,nmt} and speech recognition \cite{trojanmodel}.

Clean-label poisoning attacks were initially developed as potent threats against image classification \cite{clean-label}, and have been later adapted to the audio classification utilizing hybrid models \cite{venowave}. However, to the best of our knowledge, clean-label attacks against neural networks designed for sequence tasks have not received the same attention. Unlike attacks that rely on perceptual misalignment between input and training labels, these attacks produce poisoned samples that are hard to detect through human inspection alone \cite{hard_for_human}.
In light of this, we embark on a study to investigate the potential vulnerability of a class of state-of-the-art speech models, namely RNN-Transducers \cite{pruned}, against clean-label poisoning attacks.

Studies have also proposed defenses against clean-label attacks. These defenses can be broadly classified into three categories. The first category is a filter-type defense that aims to detect the poisoned data and remove it from the train set. 
For example, against image classification attacks, ~\cite{peri2020deep} proposes using $k$ nearest neighbors in feature space to detect poisoned samples and remove them from the training set; while~\cite{hammoudeh2021simple} proposes \textit{COSIN}-a cosine similarity influence estimator to detect the poisons. 
\cite{pmlr-v162-yang22j} found that in the gradient space, effective poisons move away from the original class. 
Hence, dropping examples in low-density gradient regions acts as an effective filtering strategy. 
The second category of defense focuses on pre-processing to "clean" the poisons without excluding data from training. 
The third category entails making the victim model more robust. \cite{geiping2021doesn} proposes to pre-train a robust foundation model by reducing adversarial feature distance and increasing inter-class feature distance to help against poisoning attacks on downstream tasks.  

The main contributions of our work are as follows:
\begin{itemize}
    \item We adapted the clean label backdoor (CLBD) attack to a spoken language understanding (SLU) task in the audio domain. While backdoor attacks have been studied in classification tasks, this work extends them to a sequence-to-sequence transduction task.
    \item We propose {\em ranked CLBD}, an improved CLBD attack that selects samples to attack based on their distance to the source class.
    \item We compared its performance against the dirty label backdoor attack, and performed a comprehensive analysis of the robustness of the attack along several axes: percentage of eligible samples poisoned; choice of poisoned instances; loudness of the trigger (as measured by signal to noise ratio), both during training and during inference; the location where the trigger is inserted with respected to the spoken command; random seed used during training.
    \item We adapted two existing defenses against adversarial attacks in the audio domain on the strongest attacks that we crafted, and examined the degree to which they are compromised in the face of an attack that is new to this domain. 
\end{itemize}

\section{Threat Model}

In this section, we describe the datasets used, the model under attack, and the various attacks and threat models considered.

\subsection{Dataset}

For our experiments, we picked the Fluent Speech Commands dataset \cite{fluent}. The dataset consists of 30,043 short utterances, each of which is annotated with three intent frames: action (e.g., ``activate"), object (e.g., ``lights"), and location (e.g., ``kitchen"). These annotations allow us to perform ``Spoken Language Understanding (SLU)" consisting of transcribing all three intent frames given the utterance. We nonetheless feel obliged to point out that it is not ``language understanding'' in the general sense.
\subsection{Spoken Language Understanding Model}
\label{ssec:slu_model}

We designed an RNN Transducer~\cite{rnn-t} model for the SLU task, implemented using the Icefall toolkit\footnote{\url{https://github.com/k2-fsa/icefall/}}. The model consisted of a $12$-layer conformer~\cite{conformer} with a hidden dimension of $16$, and a $4$-layer RNN decoder. The model was trained to emit the value of each intent in a sequence and converged in $6$ epochs, taking roughly $6$ hours on a GTX1080 graphics card. The {\em intent error rate} of the model on the benign dataset was $0.27\%$. 

\subsection{Poisoning Attacks}

\begin{figure}
  \centering
  \includegraphics[width=\linewidth]{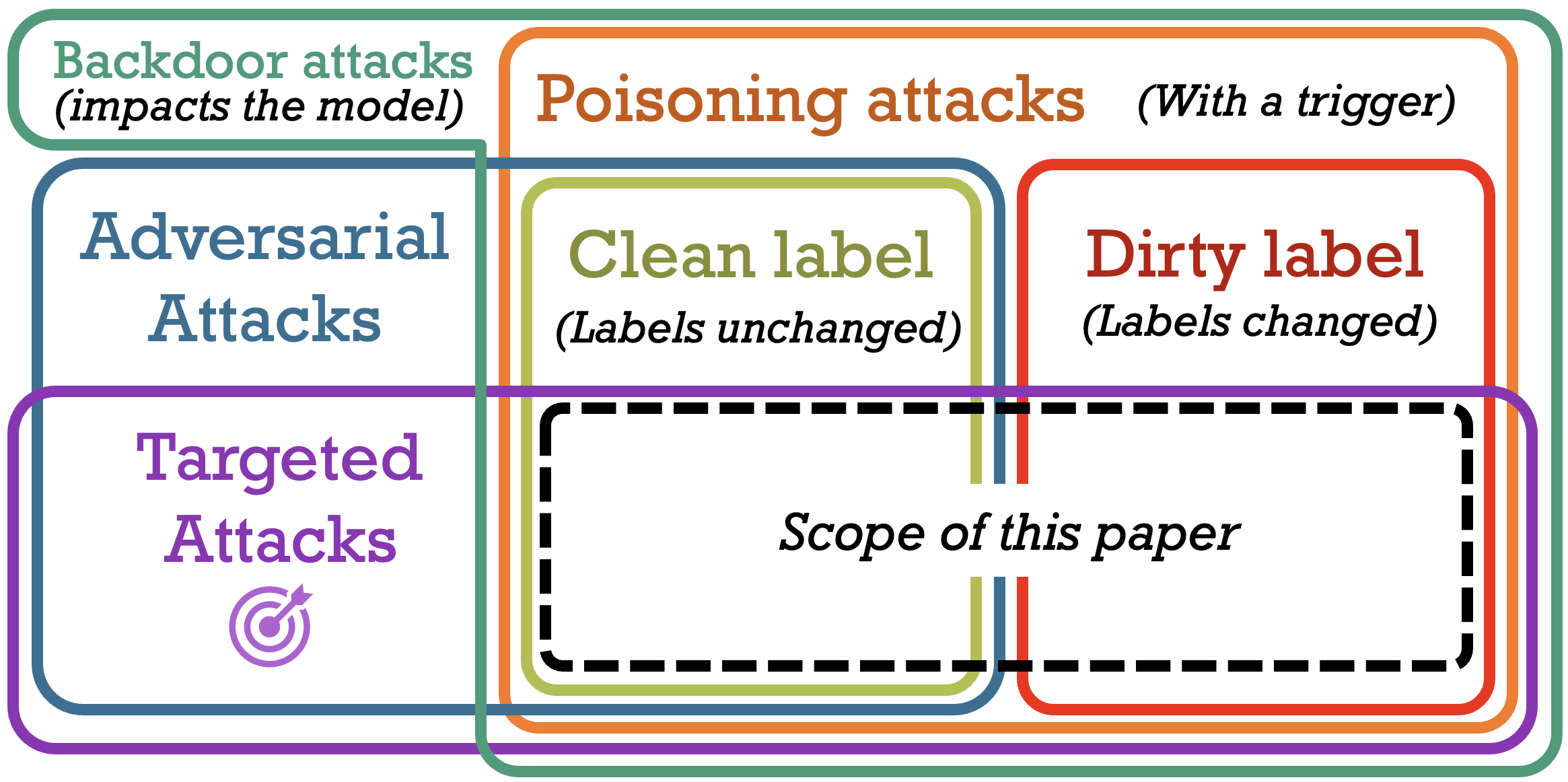}
  \caption{Classification of attacks against neural networks}
  \label{fig:poisoning_attacks}
\end{figure}
    
\begin{figure}[t]
  \centering
  \includegraphics[width=\linewidth]{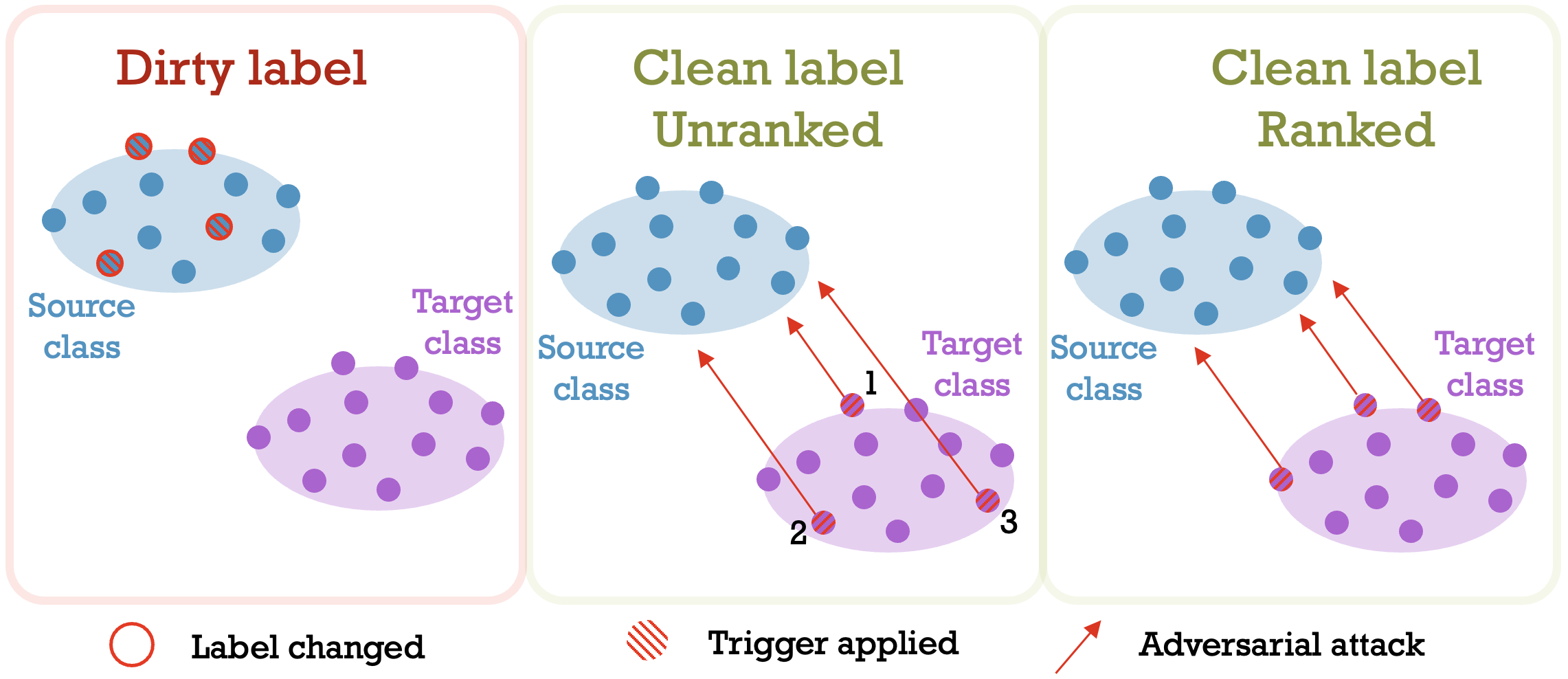}
  \caption{Illustration of dirty label, and unranked and ranked clean label poisoning attacks. In the ranked version, the poisoned utterances are selected based on the difficulty of misclassification while in the unranked version, they are selected at random.}
  \label{fig:clean_attacks}
\end{figure}

\textit{Backdoor attacks} \cite{Badnets} refer to the category of attacks where the adversary sabotages the model training to induce a malicious behavior in the presence of a trigger in the input, which the adversary may insert at will. Meanwhile, the model performs normally when the trigger is absent. \textit{Poisoning attacks} are a subclass of backdoor attacks consisting of manipulating the training data to introduce these undesired behaviors.  
Poisoning attacks are separated into two categories: \textit{Dirty label} and \textit{Clean label}. \textit{Dirty label} attacks insert a trigger on the training signal at the same time that they switch the true label (victim class) by the one the attacker wants the system to predict (target class) when the trigger is added  at inference.  Meanwhile, \textit{Clean label} attacks \textit{only} insert a trigger in the signal while keeping the original label. Therefore, \textit{Clean label} are more difficult to detect by a human observer that curates the training data.
To function properly, \textit{Clean label} attacks require combining both a trigger and an adversarial attack.
All attacks can be \textit{targeted} (changes toward a given target) or \textit{untargeted} (failure to output the correct prediction). 
Figure \ref{fig:poisoning_attacks} is a simplified diagram of the relationship between various types of attacks against neural networks. For the purpose of our study, we will be focusing on Targeted Poisoning Backdoor Attacks, both clean and dirty.
Those attacks are illustrated in Figure \ref{fig:clean_attacks}


Formally, let $M_{victim}$ refer to the victim model which will be trained on poisoned data, and let $M_{proxy}$ be a proxy model that the adversary has access to. The proxy model is assumed to be of similar architecture to $M_{victim}$ and is frozen throughout the attack process. 
$\forall s \in S_{test}$ the set of all utterances in the test set, if $c(M_{proxy}(s)) = 1$ for some eligibility criterion $c$ (which could be thought of as the indicator function of the source set), then the adversary would like to achieve the following in regard to the backdoor access function $p$ (where $p(m(s')) = 1$ indicates backdoor access for model $m$ and sample $s'$) and the trigger function $f$:

\begin{enumerate}
    \item $M_{victim}(s) = M_{proxy}(s)$: retain benign performance;
    \item $p(M_{victim}(t(s))) = 1$: insertion of trigger results in backdoor access.
\end{enumerate}



\subsubsection{Dirty Label Backdoor Attack}
\label{ssec:DLBD}

The simplest poisoning attack is the Dirty Label Backdoor Attack (DLBD) \cite{dirty}. 
In this setup, the adversary performs the following:

\begin{enumerate}
    \item Pick a subset $S_{poison}$ of the eligible utterances in the training set
    \begin{equation}
        \{s | s \in S_{train}, c(M_{proxy}(s)) = 1\}
    \end{equation}
    \item $\forall s \in S_{poison}$, replace it with $f(s)$ the same utterance with the trigger function applied;
    \item Change the training label from $l = M_{proxy}(s)$ to $l'$ (backdoor target label) such that $p(l') = 1$.
\end{enumerate}
This attack is characterized by the percentage of the source class modified (\textit{poisoning percentage}), the length, position, and strength of the trigger, as well as its nature~\cite{thebaud2023clustering, li2024multi}. Intuitively, the poisoned model associates the injected trigger with the backdoor target label.



Dirty label poisoning attacks are an example of a black-box attack: knowledge of the victim model architecture or training recipe is not required to carry out such an attack. 
Unfortunately, despite its elegant simplicity, dirty label poisoning attacks have been shown to be vulnerable to filtering-based defenses~\cite{thebaud2023clustering}.
After all, it relies on a mismatch between the salient input features in a ``human-like" perceptual space and the training label \cite{clean-label}. 

\subsubsection{Clean Label Backdoor Attack}
\label{ssec:CLBD}
Clean label backdoor (CLBD) attacks~\cite{clean-label} were thus introduced to address this very weakness. 
In a CLBD attack, the labels are unchanged, and only the signals of the target class are tampered with. The adversary carries out such an attack with the following steps:

\begin{enumerate}
    \item Pick a subset $S_{poison}$ of utterances in the training set that \textbf{satisfies the backdoor access function}: 
    \begin{equation}
        \{s|s \in S_{train}, p(M_{proxy}(s)) = 1\}
    \end{equation} with respect to the \textbf{proxy model};
    \item \textbf{$\forall s \in S_{poison}$, adversarially perturb $s$ to $s'$ so that $c(M_{proxy}(s')) = 1$} (make them look like the source class according to the proxy model).
    \item Apply the trigger function $f$ on each adversarially perturbed input $s'$.
\end{enumerate}

Intuitively, this makes the victim model perceive the attacked utterances as part of the source class, but with a trigger and a label from the target class.
Because the attack is performed before training the victim model, the adversarial attack cannot be computed on the victim model. As such, we suppose the attacker has access to the architecture used, and thus can produce a proxy model trained on the benign data. 
In this article, we explore variations in both the sample selection (1) and the trigger chosen (3), and their impact on the performances of the attack.


A crucial assumption for the success of CLBD is the transferability of adversarial perturbations from the proxy model to the victim model. The extent to which this assumption holds has been extensively studied empirically~\cite{transferability,improve_transferability}. It appears that models with similar architectures and trained on similar data generally allow cross-model transfer of adversarial samples \cite{alexa}. In order to study the most realistic attack setup possible, we allowed the attacker to be aware of the model architecture and training recipe of the victim model, but not the random seed used for training. 

The method which we use to perform adversarial perturbation in order to construct poisoned samples is known as \textbf{Projected Gradient Descent} (PGD) \cite{adversarial}, an iterative optimization method where gradient descent is performed with respect to the target label on the input while also projected into a norm-bounded box in order to not allow the input to shift too far on the perceptual space. Putting the formulation from \cite{convex} within the context of adversarial attacks, we give each iterative step of the PGD algorithm as follows:

\begin{enumerate}
    \item Compute the gradient $\nabla M_{proxy}(s_k)$ of the proxy model $M_{proxy}$ at input $s_k$ at current step $k$ with respect to the source class.
    \item Find the projected gradient $\Delta s_{pg}$ by performing a Euclidean projection on the negative gradient into the domain of $s$. As the domain of $s$ in PGD attacks is chosen to be an $l2$-norm-bounded ball with respect to a fixed center $s_0$, it is convex and therefore the aforementioned Euclidean projection is well-defined. Let the radius of the bounding ball be $r$ and the step size be denoted $\epsilon$, then the PGD update can be given by:
\end{enumerate}
\begin{equation}
\resizebox{\linewidth}{!}{
        $\Delta s_{pg} = argmin_{\Delta s_{pg}: ||\Delta s_{pg} + s_k - s_0||_2^2 < r} ||\nabla M_{proxy}(s_k) - \Delta s_{pg}||_2$ 
}
\vspace{-4mm}
\end{equation}

\subsubsection{Ranked Clean Label Backdoor Attack} 
\label{ssec:ranked}

In the original CLBD attack, the set of utterances to poison is selected at random from the target class. 
Because the adversarial perturbation is performed in order to impersonate the input as a member of the source class prior to the application of the trigger, we noted that not all eligible training samples are made equal for this purpose.
As illustrated in Figure~\ref{fig:clean_attacks}, given the same adversarial perturbation budget (as defined by some loudness threshold), some samples (samples $1$ and $2$) are more easily perturbed to look like the source class than others (such as sample $3$). 
In our experiments, we use the training loss of an utterance $s$ passed through the proxy model $M_{proxy}$ with respect to its true label as the ranking metric used in determining the relative difficulty of correctly predicting the true label of $s$: the higher the loss, the more difficult $s$ is for the proxy model, and therefore the easier it would be to adversarially perturb $s$ into something other than its true label.
In fact, our benign model mislabeled as ``activate" roughly $1\%$ of all commands in the train set that had the ``deactivate" label, and we deduced that those samples would be prime targets for applying CLBD.
We note that a very similar idea was explored in a concurrent work \cite{ranked}.

A number of other difficulty metrics may be used to rank samples by their ease of being adversarially perturbed, including the number of PGD steps or the clustering behavior of samples. We leave the comparison of these possible alternative difficulty metrics to future work.

\vspace{-3mm}
\subsection{Metrics}

We used two metrics to measure the impact of our attacks: Intent Frame Error (IFER) Rate and Attack Success Rate (ASR) ($r$).
Intent Frame Error Rate measures the percentage of intent frames that were mislabeled by the system. When trained on benign data, our system's performance fluctuates between $0.27\%$ up to $1.5\%$ IFER. 
The benign performance of the attacked systems reported in this paper falls within this range unless otherwise specified.




Attack Success Rate measures the fraction of eligible test samples that, after inserting the trigger, are predicted as the target class. 
Within the context of the task at hand, we measure the percentage of test samples with the benign label ``activate" on its ``action" intent frame that is predicted as ``deactivate", when the trigger \emph{short car horn} is additively applied.
We choose this trigger both because it is a real life noise, which would contribute to the realism of the attack, and because it was part of the set of triggers proposed in the Adversarial Robustness Toolbox (ART) in which our attack was added\footnote{\url{https://github.com/Trusted-AI/adversarial-robustness-toolbox/wiki/ART-Attacks#1-evasion-attacks}}.

\vspace{-4mm}
\section{Experiments}
In this section, we describe the experiments performed, the implementation of the dirty, clean label, and ranked clean label attacks, as well as additional analysis on the latest and two out-of-the-shelf defenses.

\subsection{DLBD and CLBD attacks configuration}
First, we compare the DLBD and CLBD attacks on the fluent speech command dataset~\cite{fluent}, using the victim model described in Section~\ref{ssec:slu_model}.
For the clean label attack, the attacker had access to a proxy model, one that was trained on the unmodified dataset and which had the same architecture as the potential victim model.
We evaluated the Attack Success Rate of the attack for poisoning percentages between 1\% and 50\%, using a \emph{short car horn} as trigger.
Then, we compared the impact of various selection strategies for the utterances to poison: selecting at random, or choosing samples that are most or least likely to be mislabeled as part of the source class.
To further analyze the ranked CLBD attack, we study the impact of various characteristics of the attack:
\begin{enumerate}
    \item Variations in the trigger strength, going from 20 to 50dB.
    \item Variations in the position of the trigger, at the start or the end of the utterance, or at random, for various percentages of poisoning.
    \item Stability of the attack, by running multiple instance of the attack using the same parameters.
\end{enumerate}

We have released the source code used in our experiments for reproducibility \footnote{\url{https://anonymous.4open.science/r/icefall_clbd-32E3/egs/slu/README.md}}.

\subsection{Proposed Defenses}
Given that CLBD attacks use gradient-based adversarial perturbations before poisoning samples, we look at two previously proposed  defenses against gradient-based attacks\footnote{
Note that neither of these defenses was trained on the Fluent Speech Commands dataset.  Their effectiveness as defenses may be reduced by domain mismatches and could potentially be restored by training or fine-tuning on the Fluent Speech Commands dataset.} \cite{joshi2024unraveling,joshi2022defense}.
Also note that both defenses were applied against a non-adaptive attack: the attacker has no knowledge that any of those could be applied.

The first is a filtering defense, which uses a binary classifier to detect samples containing adversarial noise. 
Following~\cite{joshi2024unraveling}, we train a LightResNet34 to detect adversarial noise on a dataset with 8 different adversarial attacks generated using VoxCeleb2 \cite{chung2018voxceleb2}. On an in-domain test set, this classifier reportedly achieved 98\% accuracy in distinguishing between adversarial and benign samples. We use it to automatically remove poisoned samples before training the victim model.

The second defense attempts to denoise the utterances i.e remove any adversarial noise. 
We take the denoiser from the best-performing system in ~\cite{joshi2022defense}, in which the authors adversarially fine-tuned their automatic speech recognition model and denoiser models jointly on the LibriSpeech dataset ~\cite{panayotov2015librispeech}.  They showed that this defense successfully achieved a low ground-truth word error rate (WER) and high target WER against adversarial attacks, with a very slight degradation in benign WER.   We apply it to denoise all the data before training the model.
\begin{table}
  \centering
  \caption{Dirty Label Backdoor Attack (DLBD) vs Clean Label Backdoor Attack (CLBD) results at various poisoning percentages. The trigger strength during training and inference is $20$dB. The loudness limit of the adversarial perturbation is $30$dB.}
  \label{tab:dirty}
  \begin{tabular}{ c c c }
    \toprule
    \textbf{Poisoning } &
    \multicolumn{2}{c}{\textbf{Attack Success Rate (\%)}}  \\
    \textbf{Percentage} & \textbf{DLBD} &
    \textbf{CLBD}  \\
    \midrule
        1\% & 81.7 & 1.8 \\
        2\% & 99.5 & 0.8 \\
        5\% & 94.7 & 21.0 \\
        10\% & 99.7 & 97.0 \\
        20\% & 97.6 & 94.3 \\
        50\% & 98.7 & 100 \\
    \bottomrule
  \end{tabular}
\end{table}

\section{Results and Discussions}
\subsection{DLBD versus CLBD}

Table \ref{tab:dirty} compares the Attack Success Rate between DLBD and basic CLBD attacks. Both attacks were successfully applied on our SLU task, with DLBD requiring a smaller poisoning percentage to reach an attack success rate of over $90\%$.



\begin{table}
  \centering
  \caption{Comparing the effects of the choice of poisoned instances on CLBD w.r.t. to the source class. The trigger strength during training and inference is $20$dB.}
  \label{tab:ranked_clean_results}
  \begin{tabular}{ c c c c }
    \toprule
    \textbf{Poisoning} & \multicolumn{3}{c}{\textbf{Attack Success Rate (\% )}} \\
    \cmidrule{2-4}
    \textbf{Percentage} &
    \textbf{Closer to} &
    \textbf{Random} &
    \textbf{Further to} \\
    & \textbf{Source} & & \textbf{Source} \\
    \midrule
        5\% & 79.5 & 21.0 & 35.1 \\
        10\% & 99.8 & 97.0 & 15.5 \\
        20\% & 99.7 & 94.3 & 93.2 \\
        50\% & 100.0 & 100 & 99.2 \\
    \bottomrule
  \end{tabular}
\end{table}

\subsection{Ranked CLBD}

Following on our discussion in Section \ref{ssec:ranked}, we present the attack success rate of our proposed ranked CLBD attack. In addition to comparing it against the original CLBD attack, where the poisoned samples were selected at random, we further experimented with a ``reverse-ranked CLBD attack", where the samples from the target class that are further from the source class were selected for poisoning. The results aligned with our expectations: poisoning the samples closer to the source class improves the performance of the attack compared to selecting the further ones, or doing it randomly. Using ranked CLBD, we were able to achieve an attack success rate of $99.3\%$ with just $1.5\%$ of eligible samples poisoned.

\begin{table}
  \centering
  \caption{The effects of different trigger strengths on ranked-choice CLBD. In particular, we look at the effects of using a different trigger loudness at test time compared to training time. Poisoning percentage is fixed at 10\%.}
  \label{tab:db}
  \begin{tabular}{ c c c c }
    \toprule
    \textbf{db:} &
    \multicolumn{3}{c}{\textbf{Attack Success Rate(\%)}} \\
    \textbf{Test} & 
    \textbf{Train 20dB} &
    \textbf{Train 30dB} &
    \textbf{Train 40dB} \\
    \midrule
        20 & 99.8\% & 99.5\% & 5.9\% \\
        30 & 99.4\% & 90.5\% & 4.5\% \\
        40 & 67.0\% & 32.4\% & 3.8\% \\
        50 & 10.9\% & 1.2\% & 2.9\% \\
    \bottomrule
  \end{tabular}
\end{table}

\subsection{Variations of the Ranked CLBD Attack}

\begin{table*}[th]
  \centering
  \caption{Results of our baseline defenses on ranked CLBD, as well as perfect defenses for comparison, and AUC for the binary classification task of filtering benign versus poisoned samples. ASR stands for \textit{Attack Success Rate}. As a comparison, we show the performances with an oracle filtering, that with present as "\textit{Perfect Filtering}".}
  \label{tab:defense}
  \begin{tabular}{ c c c c c c}
    \toprule
    \textbf{Poisoning} &
    \textbf{Undefended} &
    \multicolumn{2}{c}{\textbf{Filtering Defense}} &
    \textbf{Perfect Filtering} &
    \textbf{Denoiser Defense} \\
    \cmidrule{2-6}
    \textbf{Percentage} & \textbf{ASR(\%)}& \textbf{ASR(\%)} &
    \textbf{AUC} & \textbf{ASR(\%)} & \textbf{ASR(\%)} \\
    \midrule
        0\% & 1.3      & -        & -      & -    & -\\
        5\% & 79.5      & 0.9     & 0.775  & 1.2    & 4.7\\
        10\% & 99.8     & 1.1     & 0.678  & 1.5    & 82.1  \\
        20\% & 99.7     & 23.7     & 0.662  & 5.0    & 87.3\\
        50\% & 100.0    & 17.0     & 0.700  & 4.5    & 58.5\\
    \bottomrule
  \end{tabular}
\end{table*}

\subsubsection{Trigger Strength}

As our trigger function consists of additively applying a non-speech noise on top of the original utterance, reducing the strength of the applied trigger is desirable for the attacker in order to make the attack more discrete. We therefore investigated the extent to which the trigger may be weakened before we see a dropoff in attack success rate. Our results are shown in Table \ref{tab:db}. We witnessed a sharp decline in attack success rate when either the training-time or the inference-time trigger strength dropped below $30$dB.

\begin{table}
  \centering
  \caption{Comparing different trigger insertion locations for ranked-choice CLBD, across a range of poisoning percentages.}
  \label{tab:trigger_location}
  \begin{tabular}{ c c c c }
    \toprule
    \textbf{Poisoning} & \multicolumn{3}{c}{\textbf{Attack Success Rate(\%)}} \\
    \cmidrule{2-4}
    \textbf{Percentage} &
    \textbf{Start} &
    \textbf{Random} &
    \textbf{End} \\
    \midrule
        1\% & 80.8 & 2.3 & 5.6 \\
        2\% & 84.3 & 47.4 & 2.1 \\
        5\% & 79.5 & 5.1 & 3.5 \\
        10\% & 99.8 & 94.7 & 98.8 \\
        20\% & 99.7 & 6.0 & 18.9 \\
        50\% & 100.0 & 57.3 & 17.4 \\
    \bottomrule
  \end{tabular}
\end{table}

\subsubsection{Trigger Location}

The results of the insertion of the trigger in various positions are presented in Table \ref{tab:trigger_location}. 
We can see that the best results are obtained when the trigger is always at the same location, at the start of the utterance, matching the placement of the attacked intent frame within the sequence.

\subsubsection{Stability}

Over the course of our experiments, we observed that for CLBD, the increase in Attack Success Rate is highly non-linear with respect to the increase in poisoning percentage. This phenomenon is especially pronounced at very low poisoning percentages. Across the range of poisoning percentages from $0$ to just below $2.5\%$, the attack success rate jumps between less than $25\%$ and close to $100\%$. We hypothesized that the source of this non-linearity is the inherent randomness in the training of the victim model (on aspects such as weight initialization, batching, dropoff, etc.), which results in varied levels of adversarial sample transferability across the models trained in different runs. 

In order to test our hypothesis, we chose our best low-poisoning-percentage attack ($1.5\%$ poisoned, attack success rate $=99.3\%$) and repeated the attack $10$ times, each time with a different random seed. Attack success rates over those $10$ runs averaged $71.7\%$, with a standard deviation of $28.3\%$, revealing the instability of the attack in the face of random seed perturbations, just as earlier works on adversarial robustness had predicted \cite{random}. The instability is greatly mitigated at higher poisoning percentages: over a course of $10$ runs at $10\%$ poisoning percentage, we recorded an averaged attack success rate as high as $93.5\%$ and a standard deviation of just $4.4\%$. In fact, the lowest attack success rate across those $10$ runs was $87.9\%$.



\subsection{Performances of the Proposed Defenses}

The performance of the proposed defenses against the ranked CLBD attack are shown in Table \ref{tab:defense}.
The denoising defense substantially reduced attack success rate across all the poisoning percentages that we tested. We next repeated the denoising-defended $10\%$ poisoning experiment with $10$ different random seeds, observing a mean attack success rate of $46.9\%$, substantially lower than in the undefended case. 
However, the filtering defense was much more successful at producing a substantial drop in attack success rate compared to the undefended case, with an Attack Success Rate very close to the one of a perfect filter.
We think the difference is due to the nature of the defense, one removing only the adversarial noise, leaving behind the trigger for training, while the second exclude the whole file.

\section{Conclusion}

We applied clean label poisoning backdoor attacks to an RNN-T model designed for an SLU task and compared their effectiveness against dirty label poisoning backdoor attacks on the same task. 
We discovered that training samples in the target class that are closer to the source class results in the most potent attack, allowing us to achieve $99.3\%$ attack success rate with just $1.5\%$ of the eligible samples poisoned. 
We examined the effectiveness of our attack along a multitude of dimensions: loudness of the trigger, where we found the lowest loudness threshold necessary for mounting a successful attack is $30$dB; percentage of eligible samples poisoned, where we observed a highly non-linear correlation between attack success rate and poisoning percentage; location of trigger insertion, where we found that inserting the trigger in the front of utterances (corresponding to the location of the target intent frame) results in the best attack success rate. While investigating the non-linear and fluctuating correlation between attack success rate and poisoning percentage, we discovered that CLBD at low poisoning percentage is unstable in the face of random variations in the model training. 

We tested two off-the-shelf defenses trained on out-of-domain datasets and calibrated for this attack: one based on filtering and the other on pre-processing denoising, and found that the filtering-based defense was more effective, making the ASR drop to 1\% in low poisoning scenarios. 
We hypothesize that training these defenses on an in-domain dataset could enhance their effectiveness even further.
The denoiser used for the defense was trained to remove only adversarial noise, hence leaving both the original audio and the trigger behind; another direction for improved defenses would be to use a denoiser that would not only focus on adversarial attacks, but also remove natural noises.
On the other hand, in the current experimental setup, the adversary is not aware of the defenses. Designing the poison with a proxy defense in mind may make the attacks more successful even against the defended SLU system. It would thus be interesting to explore further after strengthening both the attack and defense strategies.

\bibliographystyle{IEEEtran}
\bibliography{mybib}

\end{document}